\documentclass[sigconf]{acmart}

\usepackage{multirow}
\usepackage{subfig}
\usepackage{bm}
\usepackage{tabularx}   
\usepackage[dvipsnames]{xcolor}
\usepackage{xcolor}

\AtBeginDocument{%
  }

\usepackage{multirow}
\usepackage{soul}
\newcommand{\longname}{\textit{TRend and explAnation Integrated Learner}}
\newcommand{\shortname}{\textit{TRAIL}}

\begin{document}

\title{Following the TRAIL: Predicting and Explaining Tomorrow’s Hits with a Fine-Tuned LLM}


\author{Yinan Zhang}
\authornote{Both authors contributed equally to the paper}
\orcid{0000-0002-7177-7410}
\affiliation{%
  \institution{Nanyang Technological University}
  \country{Singapore}}
\email{yinan.zhang@ntu.edu.sg}

\author{Zhixi Chen}
\authornotemark[1]
\orcid{0009-0009-5216-8722}
\affiliation{%
  \institution{Nanyang Technological University}
  \country{Singapore}}
\email{CHEN1872@e.ntu.edu.sg}

\author{Jiazheng Jing}
\orcid{0000-0002-3371-8196}
\affiliation{%
  \institution{Nanyang Technological University}
  \country{Singapore}}
\email{JIAZHENG001@e.ntu.edu.sg}

\author{Zhiqi Shen}
\orcid{0000-0001-7626-7295}
\affiliation{%
  \institution{Nanyang Technological University}
  \country{Singapore}}
\email{zqshen@ntu.edu.sg}


\begin{abstract}
  Large Language Models (LLMs) have been widely applied across multiple domains for their broad knowledge and strong reasoning capabilities. However, applying them to recommendation systems is challenging since it is hard for LLMs to extract user preferences from large, sparse user-item logs, and real-time per-user ranking over the full catalog is too time-consuming to be practical. 
  Moreover, many existing recommender systems focus solely on ranking items while overlooking explanations, which could help improve predictive accuracy and make recommendations more convincing to users.
  Inspired by recent works that achieve strong recommendation performance by forecasting near-term item popularity, we propose \shortname{} (\longname{}). \shortname{} is a fine-tuned LLM that jointly predicts short-term item popularity and generates faithful natural-language explanations. It employs contrastive learning with positive and negative pairs to align its scores and explanations with structured trend signals, yielding accurate and explainable popularity predictions. Extensive experiments show that \shortname{} outperforms strong baselines and produces coherent, well-grounded explanations.
\end{abstract}




\maketitle

\section{Introduction}

Large Language Models (LLMs) have achieved remarkable success across a variety of tasks due to the strong ability to incorporate open-world knowledge and perform complex reasoning \cite{huang2023towards, yang2025knowledge}. However, applying LLMs to recommendation systems poses two key challenges. First, while LLMs are well-suited for processing natural language, they struggle to understand user preferences from large, sparse, and structured user-item interaction data \cite{ren2024representation, Wang2025llmenhanced, yu2025application}, where users and items are typically represented by abstract ID numbers with no inherent semantic meaning. 
Second, running per-user LLM inference to score every item in a large catalog is prohibitively expensive and not practical for real-time serving \cite{zhong2024distserve}, especially on high-traffic platforms that require millisecond-level latency and tight cost constraints.

In recent years, many recommendation systems have begun to adopt popularity-based signals, especially short-term trends, as effective and efficient predictors of user engagement \cite{jing2023pare, ding2023trending, ji2020re}. By forecasting which items will become popular in the near future, the system can produce a single ranked list that is served to all users, avoiding expensive and time-consuming per-user computations. This approach is particularly effective in cold-start scenarios, where traditional recommendation methods often struggle due to their reliance on historical user interactions \cite{chaimalas2023bootstrapped}. In practice, new items such as movies or fashion products tend to attract the most attention shortly after their release, when little to no interaction data is available. By predicting near-future popularity, the system can identify and promote emerging trends early, ensuring timely and relevant recommendations even for newly launched content. 

However, most popularity-based approaches focus only on ranking and overlook the importance of explanations \cite{balloccu2022post}. Providing explanations can not only help the model make more accurate predictions by learning why an item is likely to trend, but also make the recommendations more transparent and trustworthy \cite{zhang2020explainable, peng2024uncertainty}.

Motivated by these considerations, and inspired by recent evidence that forecasting short-term item popularity can match or surpass sophisticated personalized models for top-N recommendation\cite{jing2023pare}, we revisit popularity-based recommendation through the lens of explainability. We propose \longname{} (\shortname{}), a fine-tuned LLM that jointly predicts short-term item popularity and generates natural-language explanations.

A key challenge is the absence of ground-truth explanations for why an item trends during a specific period. To address this, we construct positive and negative sample pairs based on similarity in popularity trends and item-side metadata. We then apply contrastive learning to guide the model to generate similar explanations for similar pairs. This contrastive objective not only helps the model generate more meaningful and consistent explanations, but also enhances the accuracy of popularity prediction by reinforcing consistency across items with similar temporal and contextual patterns. Extensive experiments on real-world datasets demonstrate our \shortname{} outperforms strong baseline methods.

In summary, we made the following contributions:
\begin{itemize}
    \item Motivated by LLMs’ strong reasoning capability and the success of short-term popularity prediction for top-N recommendation, we fine-tune an LLM to forecast near-future item popularity and serve one ranked list to all users. This design leverages the LLM’s language understanding to provide scalable and interpretable recommendations without requiring personalized inference for each user.
    \item To make recommendations both accurate and convincing, we pair each prediction with a natural-language explanation and train the model using contrastive learning on trend- and item-based positive/negative pairs. To our knowledge, this is the first explainable recommendation model designed specifically for popularity prediction.
    \item Extensive experiments show that \shortname{} outperforms strong baselines, and ablation studies together with human evaluations demonstrate that its explanations not only improve recommendation accuracy, but also align well with observed trends and item-side features, making them clear and convincing to users.
\end{itemize}

\section{Related Works}

\subsection{Non-Personalized Recommendation}

Non-personalized recommendation systems, which rank items for all users based solely on item-side signals, have re-emerged as scalable, interpretable, and privacy-preserving alternatives to personalized models \cite{adomavicius2005survey, ricci2011introduction}. Early approaches relied on global popularity, with studies showing that simple ``Most Popular'' methods can rival complex algorithms under sparse data \cite{herlocker2004evaluating}. Later works introduced temporal dynamics to capture short-term trends, such as time-filtered popularity and recency weighting, which significantly improved relevance \cite{phelan2009twitter, koren2010temporal, ji2020re}.

Subsequent research moved beyond pure popularity toward content- and hybrid-based item modeling. Embedding methods such as Okura et al. \cite{okura2017embeddingnews} and DKN \cite{wang2018dkn} enriched item representations with textual and knowledge-graph semantics, while UniSRec \cite{hou2022unisrec} and RecFormer \cite{li2023recformer} leveraged pretrained language models for zero-/few-shot and text-based recommendation. The recent PARE model \cite{jing2023pare} further advanced this line by forecasting near-future item popularity through multi-scale temporal attention.

Together, these developments show a clear shift from static popularity toward semantically rich and temporally adaptive item-side ranking. Given LLMs’ strong reasoning and understanding abilities, applying them to non-personalized recommendation, particularly for trend prediction, represents a promising next step for building efficient and explainable recommender systems.

\subsection{LLM-based Recommendation}

Recent advances in large language models (LLMs) have driven a surge of personalized recommendation methods aiming to model each user’s preferences and predict their ranking over all items. Early works such as P5 \cite{geng2022p5} and PEPLER \cite{li2023personalized} introduced the ``Recommendation as Language Processing'' paradigm by expressing user–item interactions in natural language. Building on this, hybrid models like LLM2BERT4Rec \cite{harte2023llm4rec}, LLM-ESR \cite{liu2025llmesr}, and GPT4Rec \cite{li2023gpt4rec} integrated LLMs with sequential recommenders through embedding alignment, summarization, or generative query modeling. More recent studies incorporated reasoning and autonomy, with agent-based systems such as RecMind \cite{wang2024recmind}, RecAgent \cite{wang2025usersim}, and InteRecAgent \cite{huang2025interecagent} treating LLMs as intelligent planners that interact with traditional recommenders for context-aware and conversational recommendations. Fine-tuned models like CoLLM \cite{zhang2025collm}, BinLLM \cite{zhang2024binllm}, and SLIM \cite{wang2024slim} further infused collaborative signals or distilled reasoning from large models. Generative paradigms such as GenRec, Chat-Rec, and LlamaRec \cite{ji2024genrec, gao2023chatrec, yue2023llamarec} extended this trend by directly generating or ranking items through textual generation and instruction tuning. Together, these works trace an evolution from prompt-based zero-shot inference to reasoning-driven and generative LLM recommenders.

While these methods improve personalization and interpretability, they face two main challenges. First, predicting a score for every user–item pair causes high inference costs, making real-time deployment impractical \cite{covington2016youtube, sun2019bert4rec, liao2022impact}. Second, relying on user and item IDs as meaningless tokens limits LLMs from using their language understanding abilities \cite{li2023recformer, hou2022unisrec, wu2024survey}.

\subsection{Explainable Recommendation}


Explainable recommendation has gained increasing importance as transparency and trust become central to recommender systems. Early studies focused on intrinsic interpretability, where factorization-based methods \cite{mcauley2013hidden, zhang2014efm, tao2019fact} linked latent factors to explicit attributes or topics to generate template-based explanations. Multimodal models such as NARRE \cite{chen2018narre} further improved interpretability by highlighting influential reviews or modalities. Later, knowledge-graph-based models including RippleNet \cite{wang2018ripplenet}, KPRN \cite{wang2019kprn}, KGCN \cite{wang2019kgcn}, and RuleRec \cite{ma2019rulerec} provided structured reasoning paths as explanations. However, these approaches typically visualize subgraphs or rules rather than producing natural-language justifications, limiting their accessibility to end users.

Recently, the rise of LLMs has opened new opportunities for generating fluent, human-readable explanations grounded in reasoning and evidence. Models such as XRec \cite{ma2024xrec}, CIER \cite{liu2025cier}, G-Refer \cite{li2025grefer}, and LANE \cite{zhao2024lane} integrate LLMs with recommender backbones to produce coherent and context-aware explanations. User studies further show that LLM-generated explanations are perceived as more natural, convincing, and helpful \cite{lubos2024llmexp}. These advances highlight the promise of LLMs in bridging accurate recommendation with trustworthy, easy-to-understand explanations.

\section{Problem Definition}

In this work, we study the task of explainable short-term popularity prediction for non-personalized recommendation. The goal is to rank items based on their expected popularity in the near future, while also generating explanations that clarify why each item is likely to trend. This enables scalable recommendation without relying on per-user inference and enhances transparency by providing interpretable predictions.

More specifically, for each item $i \in \mathcal{I}$ at time $T$, we are given a set of features including static descriptive text $d_i$, and historical trend information represented by $h_i^T = [p_i^1, p_i^2, ..., p_i^{T - 1}]$, where $p_i^t \in \mathbb{N}$ denotes the number of users who interacted with item $i$ at time $t$. In particular, $p_i^1$, corresponds to the interactions when the item was first released. 
The goal is to \textbf{predict} and \textbf{explain} the near-future popularity $p_i^T$ for each item, and to rank all items based on these predictions so that the top-ranked items can be recommended to all users in a non-personalized manner.





\section{\shortname{}}

\begin{figure*}
	\centering
	\includegraphics[width=2\columnwidth]{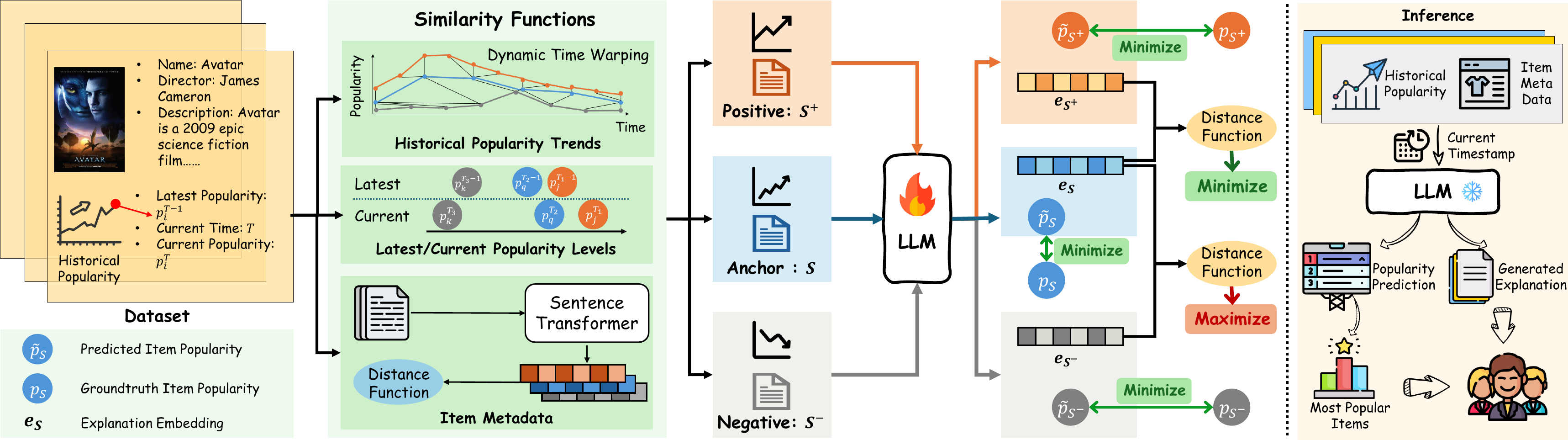}
	\caption{The architecture of \shortname{}.}
	\label{fig:arc}
\end{figure*}

As shown in Figure~\ref{fig:arc}, we propose to fine-tune an LLM to jointly predict and explain the near-future popularity of each item. A key challenges is the absence of ground-truth explanations. The training data contains popularity labels but no textual rationales describing why an item is expected to trend. To address this, we introduce a contrastive learning approach based on the assumption that samples with similar characteristics should have similar explanations.

Specifically, we assume that items with similar historical popularity trends, descriptive features, and current engagement levels are likely to trend for similar reasons. For example, if two newly released horror movies both show increasing engagement in the days near Halloween, a reasonable explanation for both might be \textit{``recently released horror film gaining traction close to Halloween"}. Though such explanations are not provided, their similar trend patterns and context suggest that a shared rationale is appropriate.

Based on this idea, we design three similarity functions to group training samples as positive (similar) or negative (dissimilar) pairs, and apply contrastive learning to bring the explanations of similar samples closer together while pushing apart those of dissimilar ones. 
In parallel, we apply a cross-entropy loss on predicted popularity scores to ensure accurate forecasting. This joint objective enables \shortname{} to generate explanations aligned with trend signals while improving prediction performance.



\subsection{Similarity Functions}
\label{sec:item_similarity}



Since ground-truth explanations are not available, we adopt a contrastive learning strategy to supervise explanation generation. The key idea is that items with similar characteristics should have similar explanations, while dissimilar items should not. Although contrastive learning is applied only to the explanation, we believe it could indirectly benefit popularity prediction through shared backbone layers by improving the underlying representations.

Motivated by recent findings in popularity-based recommendation, such as PARE \cite{jing2023pare}, which shows that popularity trends, temporal effects, and side information are key drivers of future item popularity, we define similarity from three complementary perspectives: \textit{Historical Popularity Trends}, \textit{Latest/Current Popularity Levels}, and \textit{Item Metadata}. These dimensions are selected based on both empirical evidence and intuitive reasoning \cite{jing2023pare}. Historical trends capture long-term engagement patterns. Latest or current popularity reflects short-term momentum, which has been shown to significantly influence user behavior. Item metadata, such as category, genre, or product attributes, helps explain why items attract attention, especially in cold-start scenarios. We integrate these three aspects into a unified similarity function to construct contrastive pairs, enabling the model to generate explanations that are consistent and grounded in real-world popularity signals. 


\subsubsection{Historical Trend Similarity.} 
Historical popularity trends play a critical role in understanding how items gain attention over time and are key to predicting future popularity \cite{jing2023pare, wu2017dtcn}. To incorporate this into our similarity function, it is important to compare the shape of the trend, rather than the raw interaction counts. However, modeling similarity between historical trends is challenging because items are typically released at different times, resulting in popularity sequences of varying lengths and temporal alignments.

A naive approach like cosine similarity may fail in this setting for two reasons. First, it assumes that the popularity sequences of different items are aligned in time, which is often not true since items are released at different timestamps. Second, it is sensitive to the absolute magnitude and length of the sequences. As a result, two items that follow a similar trend, such as a gradual rise or sudden drop, may be considered dissimilar simply because their sequences have different lengths or popularity levels.

To address these issues, we adopt Dynamic Time Warping (DTW) \cite{berndt1994dtw} as our similarity metric for trend sequences. DTW measures the similarity between two temporal sequences by aligning them with minimal accumulated distance, even if they are different in length. It allows flexible, non-linear alignment between sequences, enabling the model to compare the underlying trend patterns even when the sequences differ in length or contain local distortions. This is especially useful for recognizing common popularity patterns, such as rise followed by fall curves, regardless of when they occurred or how long they lasted. 
Given two sequences $h_i^{T_1}, h_j^{T_2}$, we firstly define the distance $d(\cdot, \cdot)$ for the items, $i$ and $j$ of different timestamps $t_1, t_2$ as:

\begin{equation}
    d_{i,j}(t_1, t_2) = \left|\left| p_i^{t_1} - p_j^{t_2} \right|\right|_2.
\end{equation}
Then we construct a cumulative cost matrix $D_{i,j}(t_1,t_2)$:
\begin{equation}
\begin{aligned}
    D_{i,j}(t_1,t_2) & = d_{i,j}(t_1, t_2) + 
    \\
    & \min \{ D_{i,j}(t_1 - 1, t_2), D_{i,j}(t_1, t_2-1), D_{i,j}(t_1 - 1, t_2 - 1) \}
\end{aligned}
\end{equation}
where $D(1, 1) = d(1, 1)$, $D(t, 0) = D(0, t) = +\infty$. 



Finally, the similarity between two sampled historical trends $h_i^{T_1}$ and $h_j^{T_2}$ is defined as:

\begin{equation}
\text{sim}_{\text{trend}}(h_i^{T_1}, h_j^{T_2}) = \frac{1}{1 + D_{i,j}(T_1,T_2)},
\end{equation}
where a smaller DTW distance indicates more similar trends and results in a higher similarity value.


\subsubsection{Latest/Current Popularity Levels.}
We consider both the most recent observed popularity (before the prediction window) and the ground-truth popularity during the prediction period to compute similarity. The latest popularity reflects short-term momentum, which has been shown to strongly influence future engagement \cite{anelli2019local, jing2023pare, yoo2025tpab}. Meanwhile, the ground-truth popularity provides a direct signal of actual user interest during the target period. By combining these two signals, we capture both pre-trend momentum and actual trend, offering a more complete picture of item-level similarity.

More specifically, we define the similarity between two samples, where item $i$ is observed at time $T_1$ and item $j$ at time $T_2$, as follows:
\begin{equation}
\begin{aligned}
    \text{sim}_{\text{latest}} (i, j, T_1, T_2) &= \exp \left(- \frac{(r_i^{T_1} - r_j^{T_2})^2}{2\sigma^2}\right), 
\end{aligned}
\end{equation}
where $r_i^{T_1} = \frac{p_i^{T_1} - p_i^{T_1- 1}}{p_i^{T_1 - 1}}$, which refers to the popularity change rate of item $i$ at time $T_1$. 
By comparing the distance between the latest and current popularity levels within each sample, rather than directly comparing the same type of popularity between different items, it better captures how an item's popularity is evolving over time. 


\subsubsection{Meta Data Similarity.}
While popularity trends capture how user engagement evolves over time, item metadata often plays a crucial role in why an item becomes popular. Item metadata such as category, release timing, and textual descriptions can heavily influence user interest, enhance the relevance of recommendations, and powerful in the cold-start recommendation scenario \cite{mooney2000content, choi2012genre, saveski2014coldstart, behera2022metadata}.
For instance, a newly released romantic movie just before Valentine’s Day may attract significant attention not just because of recent trends, but also due to its theme and timing. These semantic cues provide valuable context that cannot be captured by numerical popularity data alone.

To incorporate this semantic information, we leverage sentence transformers, which are well-suited for capturing contextual and semantic similarity between texts \cite{song2020mpnet}. Each item's metadata is encoded into a dense embedding using a pretrained sentence transformer, and similarity is then computed via cosine similarity between item embeddings.
This allows the model to recognize meaningful relationships such as shared genres or similar themes that are not reflected in popularity patterns, providing a complementary signal that enhances explanation generation and contrastive learning. We use a fixed sentence transformer named all-mpnet-base-v2\footnote{https://huggingface.co/sentence-transformers/all-mpnet-base-v2} to encode the description text to embedding $\mathbf{d}_i$.

The metadata similarity between item $i$ and item $j$ is defined as:
\begin{equation}
\text{sim}_{\text{meta}}(i,j)
= \cos \big(\mathbf{d}_i, \mathbf{d}_i\big), 
\end{equation}

\subsubsection{Total Similarity Function.}
To obtain the final item similarity, we integrate the three similarity functions. More specifically, for two samples, $S_1$ and $S_2$, where item $i$ is observed at time $T_1$ and item $j$ at time $T_2$ respectively, the similarity is defined as:
\begin{equation}
\begin{aligned}
\text{sim}&(S_1, S_2)  = \\ &\alpha \text{sim}_{\text{trend}}(h_i^{T_1}, h_j^{T_2}) +  \beta \text{sim}_{\text{latest}}(i,j,T_1,T_2) + 
 \gamma\text{sim}_{\text{meta}}(i,j),
\end{aligned}
\end{equation}

We use $\alpha, \beta, \gamma$ as hyperparameters to balance the contributions of the three similarity components.
For each anchor sample $S$, we select the most similar and most dissimilar samples based on the overall similarity score, and use them to construct contrastive triplets $(S, S^+, S^-)$ for training.

\subsection{Learning Objectives}

To effectively train \shortname{}, we adopt a dual-objective learning framework that aims to improve both prediction accuracy and explanation quality. First, we introduce a contrastive learning objective by learning from the contrastive triplets. This encourages the model to pull similar items closer in the explanation embedding space and push dissimilar ones further apart. Although this objective focuses on explanation generation, it also helps improve the shared representations, which can indirectly benefit popularity prediction.

In parallel, we apply a supervised loss to directly optimize popularity prediction. For item $i$ at time $T$, the model minimizes the difference between the predicted popularity score $\tilde{p}_i^T$ and the ground-truth value $p_i^T$. Together, these two objectives enable \shortname{} to make predictions that are not only accurate but also accompanied by consistent and meaningful explanations.


\subsubsection{Contrastive Loss}

To guide the generation of meaningful explanations, we assume that items with similar characteristics should have similar explanations. 
Specifically, we use the item similarity function introduced in Section~\ref{sec:item_similarity} to build the triplets. For each anchor sample $S$, we select positive samples $S^+ \in \mathcal{S^+}$ from its top-N most similar items, and negative samples, $S^- \in \mathcal{S^-}$ from the remaining less similar items. 




To improve discriminability, we apply a projection head that maps sample embedding $\mathbf{e}_S^\prime$ into a lower-dimensional space suitable for contrastive optimization. Following the design of SimCLR \cite{chen2020simclr}, we first project the explanation embeddings into a contrastive space using two lightweight non-linear projection head and layer normalization.

\begin{equation}
\begin{aligned}
    \mathbf{h}_S &= \sigma(W_1 \mathbf{e}_S^\prime + b_1), \\
    \tilde{\mathbf{h}}_S &= \mathbf{h}_S \odot m, \\
    \mathbf{e}_S &= \text{LayerNorm}(W_2 \tilde{\mathbf{h}}_S), \\
\end{aligned}
\end{equation}
where $\mathbf{e}_S^\prime$ denotes the original explanation embedding of sample $S$, $\odot$ denotes the dropout operation while $m$ stands for the random mask, and $\mathbf{e}_S$ is the projected representation. For projection head parameters, $W_1 \in \mathbb{R}^{d_{\text{hid}} \times d_{\text{in}}}$ and $W_2 \in \mathbb{R}^{d_{\text{out}} \times d_{\text{hid}}}$ are trainable weight matrices, $b_1$ is a bias term.

Given an anchor sample $S$, its positive sample set $\mathcal{S^+}$ and negative sample set $\mathcal{S^-}$, with the projected explanation embedding $\mathbf{e}_S$. We can formulate the contrastive objective with the InfoNCE loss:

\begin{equation}
    \mathcal{L}_{\text{InfoNCE} (S)} = - \log \frac{ \sum\limits_{S^+ \in \mathcal{S^+}} \exp (\mathbf{e}_S^\top \mathbf{e}_{S^+}^{}) / \tau} 
    {\sum\limits_{S^+ \in \mathcal{S^+}} \exp (\mathbf{e}_S^\top \mathbf{e}_{S^+}^{}) / \tau + \sum\limits_{S^- \in \mathcal{S^-}} \exp (\mathbf{e}_S^\top \mathbf{e}_{S^-}^{}) / \tau}
\end{equation}


This design enforces consistency in the embedding space, such that items identified as similar exhibit embeddings that are close, while dissimilar items are pushed apart. In doing so, the contrastive learning objective not only enhances the interpretability of the generated explanations, but also provides complementary signals that support the accurate prediction of item popularity scores.

\subsubsection{Supervised Loss.}
The main objective of \shortname{} is to predict the popularity score $\tilde{p}_i^T$ for each item $i$ at a target timestamp $T$. Since the LLM outputs are normalized probability distributions after the softmax layer, we employ a cross-entropy objective.
Formally, the supervised loss is defined as:






\begin{equation}
\mathcal{L}_{\text{sup}}
= - \mathbb{E} [p_i^T \log \tilde{p}_i^T],
\end{equation}
where $p_i^T$ denotes the ground-truth probability, and $\tilde{p}_i^T$ is the model’s predicted probability. The supervised loss can thus be interpreted as minimizing the divergence between these two distributions, enforcing a supervised alignment that guides the model towards more accurate popularity estimation.

\subsubsection{Overall Loss Function}
To jointly optimize our learning objectives, we define the overall loss as

\begin{equation}
    \mathcal{L} = \mathcal{L}_{\text{sup}} + \lambda \, \mathcal{L}_{\text{con}},
\end{equation}
where $\lambda$ is a hyper-parameter that controls the relative contribution of the contrastive loss. 
This joint objective ensures that \shortname{} learns to predict popularity scores accurately while also leveraging auxiliary signals from sample similarities, enhancing both prediction performance and interpretability.

\subsection{Recommendation with \shortname{}}
As shown in the right side of Figure~\ref{fig:arc}, we use \shortname{} to predict the future popularity of items and rank them accordingly. The top-ranked items, expected to trend in the near future, are then recommended to all users as the final recommendation list.

That is, for each item $i$ at target time $T$, given its historical popularity $h_i^T$, metadata description text $d_i$, and a target timestamp $T$, the model outputs both a predicted popularity score $\tilde{p}_i^T$ and a corresponding natural language explanation. We then rank all candidate items according to their predicted scores and then recommend the top-N items with the highest scores. 
The accompanying explanations improve transparency by clarifying why each item is expected to achieve that level of popularity.


\section{Experiments and Discussion}
\subsection{Experimental Settings}\label{sec:experimental_settings}

\subsubsection{Datasets}

We evaluate the proposed \shortname{} on three real-world datasets: \textit{Douban Movies}, \textit{Amazon Beauty}, and \textit{Amazon Baby}. 
The \textit{Douban Movies} dataset was collected from Douban\footnote{https://www.douban.com}, a major Chinese social media platform where users actively review and rate movies, books, music, and more. We include all movie-related user interactions that occurred between January 1, 2010, and November 30, 2019. For each movie, we construct a descriptive text by combining its name, director, brief introduction, and other relevant metadata. Note that user review content is not included.
We also evaluate our model on two public Amazon review datasets: \textit{Amazon Beauty} and \textit{Amazon Baby}. Table~\ref{tab:dataset_stats} summarizes the key statistics of all three datasets. 


\begin{table}[htbp]
\centering
\caption{Statistics of the datasets.}
\label{tab:dataset_stats}
\resizebox{0.95\linewidth}{!}{
\begin{tabular}{ccccc}
\toprule
\textbf{Dataset} & \textbf{\# Users} & \textbf{\# Items} & \textbf{\# Interactions} & \textbf{ Sparsity ($\%$)} \\
\midrule
\textit{Douban Movies}   & $33,756$ & $3,393$ & $329,579$ & $99.71 $ \\
\textit{Amazon Beauty}     & $22,126$ & $10,281$ & $159,924$ & $99.93$ \\
\textit{Amazon Baby}   & $19,280$ & $6,218$ & $135,100$ & $99.89$ \\
\bottomrule
\end{tabular}}
\end{table}

\begin{table*}[t!]
    \centering
    \caption{Model performances comparison. The best results are bold, and the second-best are underlined. The \textit{Rel. Imp.} denotes the relative improvement of \shortname{} over the best baselines.}
    \vspace{-0.4cm}
    \label{tab:main}
    \resizebox{\textwidth}{!}{
    \begin{tabular}{c|cccc|cccc|cccc}
    \toprule
\multirow{2}{*}{Methods} & \multicolumn{4}{c|}{\textit{Douban Movie}}                             & \multicolumn{4}{c|}{\textit{Amazon Baby}}                              & \multicolumn{4}{c}{\textit{Amazon Beauty}}                            \\ 
                         & HR@5            & HR@10           & NDCG@5          & NDCG@10         & HR@5            & HR@10           & NDCG@5          & NDCG@10         & HR@5            & HR@10           & NDCG@5          & NDCG@10         \\ \midrule
BPR                      & 0.0084          & 0.0114          & 0.0033          & 0.0037          & 0.0076          & 0.0190          & 0.0008          & 0.0018          & 0.0043          & 0.0117          & 0.0006          & 0.0016          \\
UserKNN                  & 0.0119          & 0.0178          & 0.0043          & 0.0055          & 0.0061          & 0.0144          & 0.0009          & 0.0024          & 0.0100          & 0.0121          & 0.0017          & 0.0017          \\
SASRec                   & 0.0346          & 0.0568          & 0.0123          & 0.0161          & 0.0190          & 0.0409          & 0.0049          & 0.0078          & 0.0173          & 0.0364          & 0.0031          & 0.0053          \\
Caser                    & 0.0257          & 0.0291          & 0.0078          & 0.0081          & 0.0159          & 0.0265          & 0.0025          & 0.0035          & 0.0113          & 0.0195          & 0.0013          & 0.0020          \\
HGN                      & 0.0321          & 0.0494          & 0.0067          & 0.0098          & \underline{ 0.0311}    & 0.0508          & 0.0079          & 0.0100          & 0.0234          & 0.0438          & 0.0038          & 0.0064          \\
Diff                     & 0.0118          & 0.0215          & 0.0062          & 0.0094          & 0.0136          & 0.0232          & 0.0050          & 0.0075          & 0.0228          & 0.0472          & 0.0035          & 0.0086          \\ \midrule
LLM2BERT4Rec             & 0.0143          & 0.0232          & 0.0039          & 0.0054          & 0.0137          & 0.0281          & 0.0066          & 0.0102          & 0.0126          & 0.0191          & 0.0037          & 0.0044          \\
LLM-ESR                  & 0.0296          & 0.0420          & 0.0099          & 0.0117          & 0.0182          & 0.0326          & 0.0031          & 0.0050          & 0.0191          & 0.0325          & 0.0021          & 0.0038          \\
LLM-TRSR                 & 0.0188          & 0.0370          & 0.0048          & 0.0078          & 0.0144          & 0.0318          & 0.0060          & 0.0093          & 0.0134          & 0.0303          & 0.0020          & 0.0034          \\\midrule
PARE                     & \underline{ 0.2222}    & \underline{ 0.2504}    & \underline{ 0.1423}    & \underline{ 0.1482}    & 0.0304          & \underline{ 0.0689}    & \underline{ 0.0094}    & \underline{ 0.0133}    & \underline{ 0.0401}    & \underline{ 0.0550}    & \underline{ 0.0088}    & \textbf{0.0121} \\
GPT-4o-mini              & 0.2099          & 0.2463          & 0.0744          & 0.0818          & 0.0192          & 0.0435          & 0.0032          & 0.0051          & 0.0302          & 0.0466          & 0.0057          & 0.0077          \\
DeepSeek                 & 0.2173          & 0.2500          & 0.1388          & 0.1441          & 0.0298          & 0.0627          & 0.0061          & 0.0098          & 0.0302          & 0.0495          & 0.0065          & 0.0097          \\ \midrule
\shortname{}             & \textbf{0.3667} & \textbf{0.4163} & \textbf{0.1970} & \textbf{0.2039} & \textbf{0.0430} & \textbf{0.0743} & \textbf{0.0123} & \textbf{0.0154} & \textbf{0.0443} & \textbf{0.0594} & \textbf{0.0088} & \underline{ 0.0108}    \\
\textit{Rel. Imp.}                & 65.05\%         & 66.25\%         & 38.44\%         & 37.56\%          & 38.45\%         & 7.81\%          & 30.75\%         & 15.94\%         & 10.61\%         & 7.95\%          & 0.89\%          & -10.34\%       \\ \bottomrule
\end{tabular}
    }
\end{table*}

To better reflect real-world recommendation scenarios, we split each dataset using a global timeline. Specifically, the entire interaction history is divided into consecutive, non-overlapping 30-day windows, where $t = 1$ represents the earliest 30-day period, $t = 2$ is the next, and so on.

This temporal splitting strategy offers two main benefits. First, it mimics real-world conditions where future interactions are unknown during training and must be predicted based solely on historical data. Second, it prevents information leakage between the training, validation, and test sets, ensuring a fair evaluation of the model’s ability to forecast future popularity trends and user behavior. Note that in this setup, items in the test set may have very limited or even no interaction history in training or validation.



\subsubsection{Baselines}

To ensure a thorough and fair comparison, we select a diverse set of baseline methods, including , personalized LLM-based approaches, and popularity-based baselines.

\noindent\textbf{Personalized Sequential Recommendation Methods:}
\begin{itemize}
    \item \textbf{BPR} \cite{rendle2009bpr}, which is a pairwise ranking method that optimizes user preference by ensuring positive items are ranked higher than sampled negatives.
    \item \textbf{UserKNN} \cite{sarwar2001itemknn}, a strong item-based collaborative filtering method that applies the k-nearest neighbors algorithm to recommend items by exploiting item–item similarities from users’ historical interactions.
    \item \textbf{SASRec} \cite{kang2018sasrec}, a self-attentive sequential recommendation model that leverages the Transformer \cite{vaswani2017attention} architecture to capture users' long-term and short-term preferences from interaction sequences.
    \item \textbf{Caser} \cite{tang2018caser}, a convolutional sequence embedding model that treats recent user interactions as an image to capture both sequential patterns and general preferences.
    \item \textbf{HGN} \cite{ma2019hgn}, a hierarchical gating network that adaptively selects relevant features and past interactions through feature- and instance-level gating, capturing both short- and long-term user preferences in sequential recommendation.
    \item \textbf{DIFF} \cite{kim2025diff}, a side-information integrated sequential recommendation model that employs frequency-based noise filtering and dual early–intermediate fusion to capture robust correlations across item IDs and attributes.
\end{itemize}

\noindent\textbf{Personalized LLM-based Recommendation Methods:}
\begin{itemize}
    \item \textbf{LLM2BERT4Rec}\footnote{https://github.com/dh-r/LLM-Sequential-Recommendation} \cite{harte2023llm4rec}, a hybrid approach that initializes the sequential recommendation model BERT4Rec \cite{sun2019bert4rec} with item embeddings obtained from an LLM.
    \item \textbf{LLM-ESR}\footnote{https://github.com/liuqidong07/LLM-ESR} \cite{liu2025llmesr}, a sequential recommendation framework that leverages semantic embeddings from LLMs to address long-tail challenges, combining dual-view modeling for item semantics with retrieval-augmented self-distillation to improve user preference representation.
    \item \textbf{LLM-TRSR}\footnote{https://github.com/zhengzhi-1997/LLM-TRSR} \cite{zheng2024llmtrsr}, a framework for text-rich sequential recommendation that segments user behaviors and applies LLM-based summarization (hierarchical and recurrent) to compress long histories, then constructs prompts for fine-tuning with SFT \cite{zhang2023instruction} and LoRA \cite{hu2022lora}.
\end{itemize}

\noindent\textbf{Popularity-based Recommendation Methods:}
\begin{itemize}
    \item \textbf{PARE} \cite{jing2023pare}, a non-personalized recommendation method that predicts future popular items by modeling popularity history, temporal and periodic impacts, and side information.
    \item \textbf{GPT-4o-mini}\footnote{\url{https://openai.com/index/gpt-4o-mini-advancing-cost-efficient-intelligence/}}, a widely used language model designed for cost-efficient inference. We use it as a recommendation baseline by prompting it to generate popularity scores for candidate items.
    \item \textbf{DeepSeek} \cite{deepseek2025r1}, a recently released large language model known for strong general-purpose reasoning. We apply it to recommendation by prompting it to predict item popularity and rank candidates accordingly.
\end{itemize}

\begin{figure*}
    \centering
    \subfloat[]{\includegraphics[trim={0cm 0cm 0cm 0cm},clip,width=.6\columnwidth]{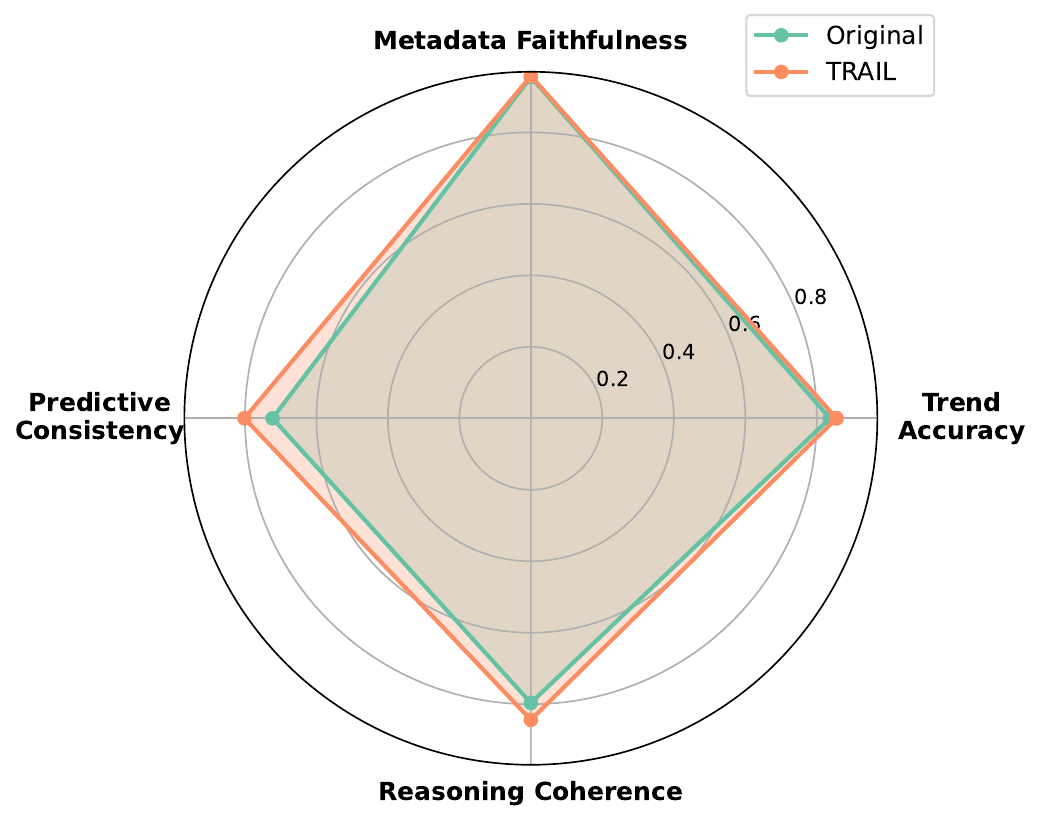}\label{fig:expain1}}
    \hfill
    \subfloat[]{\includegraphics[trim={0cm 0cm 0cm 0cm},clip,width=.6\columnwidth]{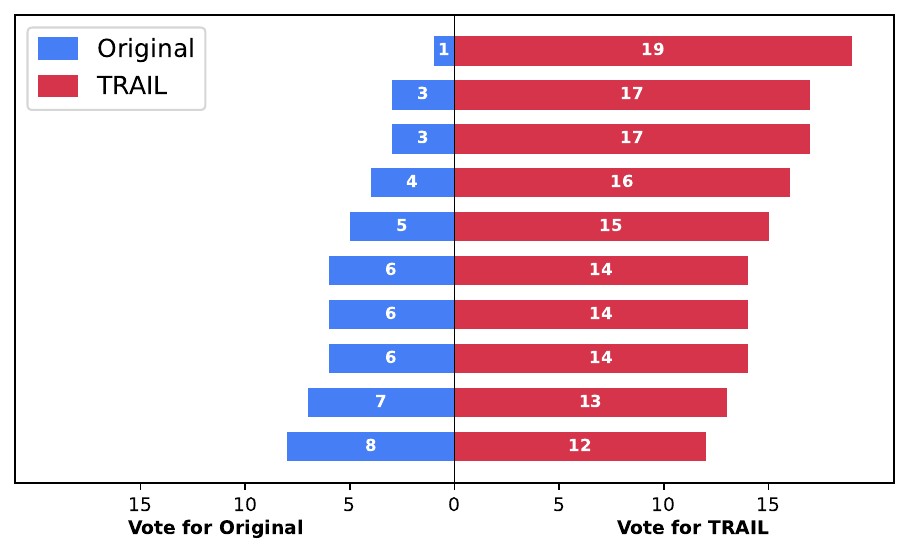}\label{fig:expain2}}
    \hfill
    \subfloat[]{\includegraphics[trim={0cm 0cm 0cm 0cm},clip,width=.57\columnwidth]{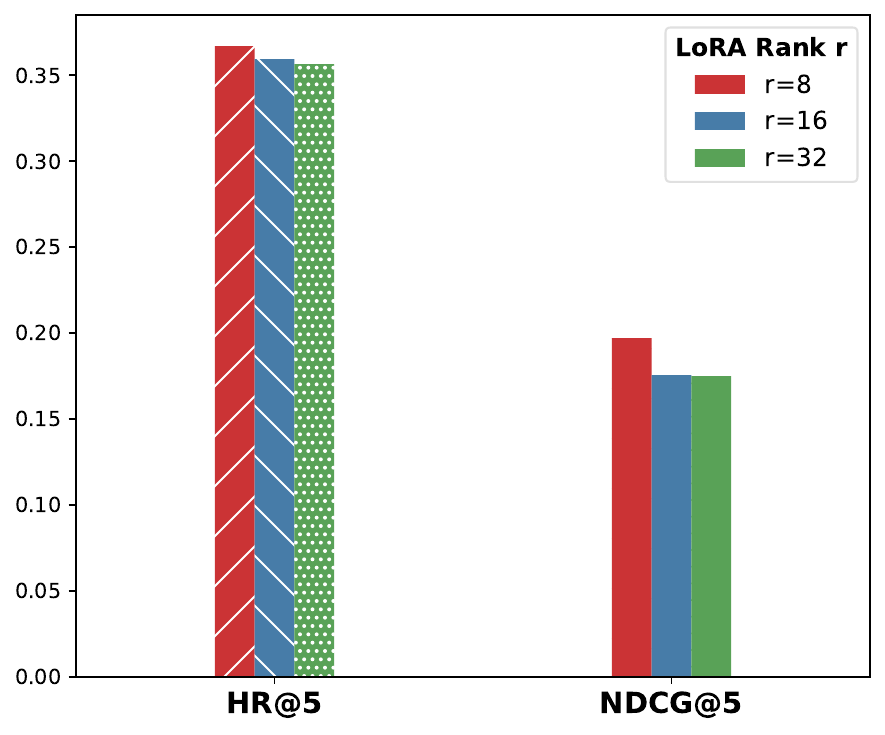}\label{fig:hyper}}
    \caption{(a) Systematic LLM-based evaluation of explanation quality for \shortname{} and the the original LLM across four dimensions. (b) Human evaluation of explanation quality for \shortname{} and the original LLM baseline without finetuning. (c) Impact of LoRA rank on recommendation accuracy.}
    \label{fig:experiment_3}
\end{figure*}

\subsubsection{Evaluation Metrics}
We evaluate \shortname{} from two key perspectives: recommendation performance and accuracy of popularity prediction. 
For recommendation performance, we use two standard ranking-based metrics. HR@K (Hit Ratio) checks whether the ground-truth popular items appear in the top K predicted list. NDCG@K (Normalized Discounted Cumulative Gain) further considers the ranking positions, assigning higher importance to correct items that appear closer to the top of the list.

To assess the accuracy of popularity prediction, we use Jaccard distance (Jaccard@K), which measures the overlap between the ground-truth top K most popular items and the model’s top K predicted items. This reflects how well the predicted popularity scores align with the actual popularity rankings.



\subsubsection{Implementation Details}

In this work, we choose DeepSeek-7B\footnote{https://huggingface.co/deepseek-ai/DeepSeek-R1-Distill-Qwen-7B} as the pretrained LLM backbone and optimize the model using the AdamW optimizer. To enable parameter-efficient fine-tuning, we apply the LoRA framework, which introduces low-rank adapter modules into the attention layers of the LLM. Each experiment is run three times, and we report the average results. Due to space limitation, additional implementation details are provided in the Appendix, and our code is publicly available\footnote{https://github.com/Revive-dontwanttocode/TRAIL}.

\subsection{Results and Discussion}


\begin{table}[t!]
    \centering
    \caption{Ablation results comparing \shortname{}, its variants, PARE, and the ground-truth upper bound on \textit{Douban Movie} and \textit{Amazon Baby}.}
    \vspace{-0.1cm}
    \label{tab:ablation}
    \resizebox{\columnwidth}{!}{
    \begin{tabular}{cc|ccc|cc}
    \toprule
Dataset                                & Metrics    & w/o $\mathcal{L}_{\text{sup}}$ & w/o $\mathcal{L}_{\text{con}}$ & \shortname{} & PARE  & Groundtruth \\ \midrule
\multirow{6}{*}{\textit{\begin{tabular}[c]{@{}c@{}}Douban \\      Movie\end{tabular}}} & HR@5       & 0.1837              & 0.2546               & 0.3667 & 0.2222 & 0.5211       \\
                                       & HR@10      & 0.2019              & 0.2716               & 0.4163 & 0.2504 & 0.6254       \\
                                       & NDCG@5     & 0.1026              & 0.1176               & 0.1970 & 0.1423 & 0.2552       \\
                                       & NDCG@10    & 0.1097              & 0.1204               & 0.2039 & 0.1482 & 0.2856       \\ \cmidrule{2-7}
                                       & Jaccard@5  & 0.0000              & 0.2262               & 0.2500 & 0.1111 & -       \\
                                       & Jaccard@10 & 0.0000              & 0.1204               & 0.1949 & 0.0526 & -       \\ \midrule
\multirow{6}{*}{\textit{\begin{tabular}[c]{@{}c@{}}Amazon \\      Baby\end{tabular}}}  & HR@5       & 0.0307              & 0.0329               & 0.0430 & 0.0304 & 0.0497       \\
                                       & HR@10      & 0.0624              & 0.0652               & 0.0743 & 0.0689 & 0.0900       \\
                                       & NDCG@5     & 0.0065              & 0.0074               & 0.0123 & 0.0094 & 0.0127       \\
                                       & NDCG@10    & 0.0095              & 0.0115               & 0.0154 & 0.0133 & 0.0191       \\\cmidrule{2-7}
                                       & Jaccard@5  & 0.1111              & 0.1574               & 0.2632 & 0.2500 & -       \\
                                       & Jaccard@10 & 0.1270              & 0.1379               & 0.3968 & 0.3333 & -     \\ \bottomrule
\end{tabular}
}
\vspace{-0.2cm}
\end{table}

\subsubsection{Recommendation Performances}
As shown in Table~\ref{tab:main}, we compare \shortname{} with a diverse set of strong and state-of-the-art recommendation models across three real-world datasets.
We made the following observations. 
First, \shortname{} consistently outperforms all baselines across all evaluation metrics, except for NDCG@10 on the \textit{Amazon Baby} dataset, where it ranks second. For instance, on the \textit{Douban Movies} dataset, \shortname{} improves over the best baseline by 65.05\% in HR@5 and 38.44\% in NDCG@5.

Second, we observe that non-personalized methods based on popularity prediction perform remarkably well. Both our proposed \shortname{} and the simple baseline PARE outperform a range of strong personalized models, including traditional, deep learning-based, and LLM-based approaches. This highlights the effectiveness of forecasting item popularity for recommendation, especially in settings where personalized interaction data may be sparse or unreliable.

Finally, compared to powerful but untrained commercial LLMs such as DeepSeek and GPT-4o-mini, our fine-tuned model achieves better performance despite having fewer parameters. We also conducted additional experiments on Douban using a few-shot GPT-4o-mini baseline with five in-context examples. TRAIL still outperformed this baseline, achieving higher hit rates than GPT-4o-mini (HR@5 = 0.2121, HR@10 = 0.2858).
This highlights both the efficiency and effectiveness of our approach. 

\subsubsection{Ablation Study}

\begin{figure}[t]
    \centering
    \subfloat[]{%
        \includegraphics[width=0.48\columnwidth]{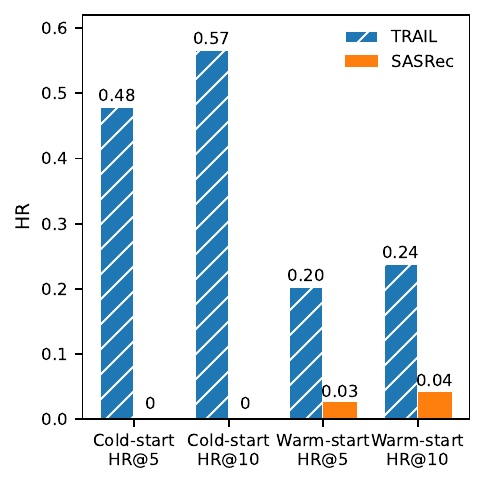}%
        \label{fig:hr_compare_sasrec}
    }
    \hfill
    \subfloat[]{%
        \includegraphics[width=0.48\columnwidth]{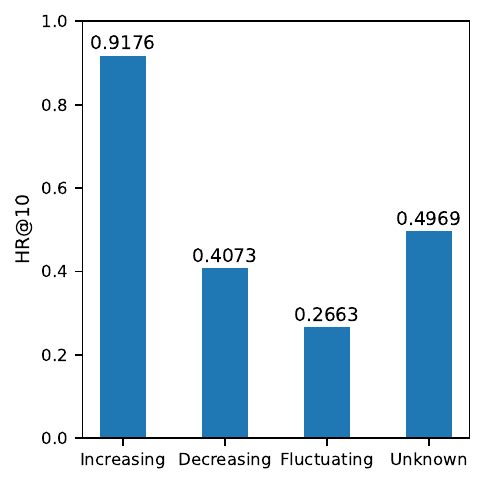}%
        \label{fig:hr_four_category}
    }
    \vspace{-0.1cm}
    \caption{(a) Performances of \shortname{} and SASRec under cold-start and warm-start settings. (b) HR@10 performance across four types of trend categories.}
    \label{fig:experiment_explain}
    \vspace{-0.4cm}
\end{figure}

\begin{table*}[t]
  \caption{Case study results on \textit{Douban Movie} and \textit{Amazon Baby}. 
  For popularity prediction, values are shown as $a(b)$, where $a$ is the predicted popularity score and $b$ is the corresponding ranking.}
  \vspace{-0.3cm}
  \resizebox{\textwidth}{!}{
  \centering
  \label{tab:case}
  \renewcommand{\arraystretch}{1.2}
  \small
  \begin{tabular}{
    >{\arraybackslash}m{0.3\linewidth} 
    >{\centering\arraybackslash}m{0.2\linewidth} 
    >{\arraybackslash}m{0.45\linewidth}
    >{\arraybackslash}m{0.04\linewidth}
    >{\arraybackslash}m{0.04\linewidth}
    >{\arraybackslash}m{0.1\linewidth}
  }
    \toprule
    \textbf{Item Description} & 
    \textbf{Historical  Trend} & 
    \textbf{\shortname{} Explanation} &
    \textbf{\shortname{}} &
    \textbf{PARE} &
    \textbf{Groundtruth} \\
    \midrule
    The movie My Dear Liar (2019) is a \textcolor{ForestGreen}{Plot, Comedy, Love} film released in 2019-11-07. In this movie, Wu Hai (played by \textcolor{ForestGreen}{Dapeng}) deliberately got acquainted with Miaomiao (played by \textcolor{ForestGreen}{Liu Yan})... It is in Mandarin, \textcolor{ForestGreen}{directed by Ao Shen} ... And \textcolor{ForestGreen}{the main actors (top-5) include Da Peng, Yan Liu, Zixian Zhang}. The movie is produced by Hao Ning and Da Peng ...
    & Just Released, {\color{red} No Historical Popularity Trend.}
    & \textbf{Trend}: \textcolor{red}{Without historical popularity data}, we cannot assess trends over time. \textbf{Feature}: The movie's genre {\color{ForestGreen} (Comedy)}, strong {\color{ForestGreen} cast featuring Da Peng, Yan Liu, and Zixian Zhang}, and {\color{ForestGreen} experienced director Ao Shen} contribute to its potential success. \textbf{Integration}: Combining the movie's genre appeal, a strong cast, and quality production with the absence of negative factors suggests a high popularity score.
    & 85 \  (5)
    & 0 (>50)
    & 154 (6)
    \\ \midrule
    The baby product name is \textcolor{ForestGreen}{Baby Banana Bendable} Training \textcolor{Green}{Toothbrush}, Infant; The \textcolor{ForestGreen}{flexible }infant baby banana brush is ... The \textcolor{ForestGreen}{fun banana shape} and \textcolor{ForestGreen}{soft texture} are the perfect ... It is made of the \textcolor{ForestGreen}{highest quality silicone} and is compliant with the highest government standards. Because of its unique \textcolor{ForestGreen}{flexibility}, it is much safer ... The \textcolor{ForestGreen}{soft silicone} can help sooth sore gums and it is \textcolor{ForestGreen}{dishwasher safe} and \textcolor{ForestGreen}{freezer friendly}......
    & 
    \includegraphics[width=\linewidth]{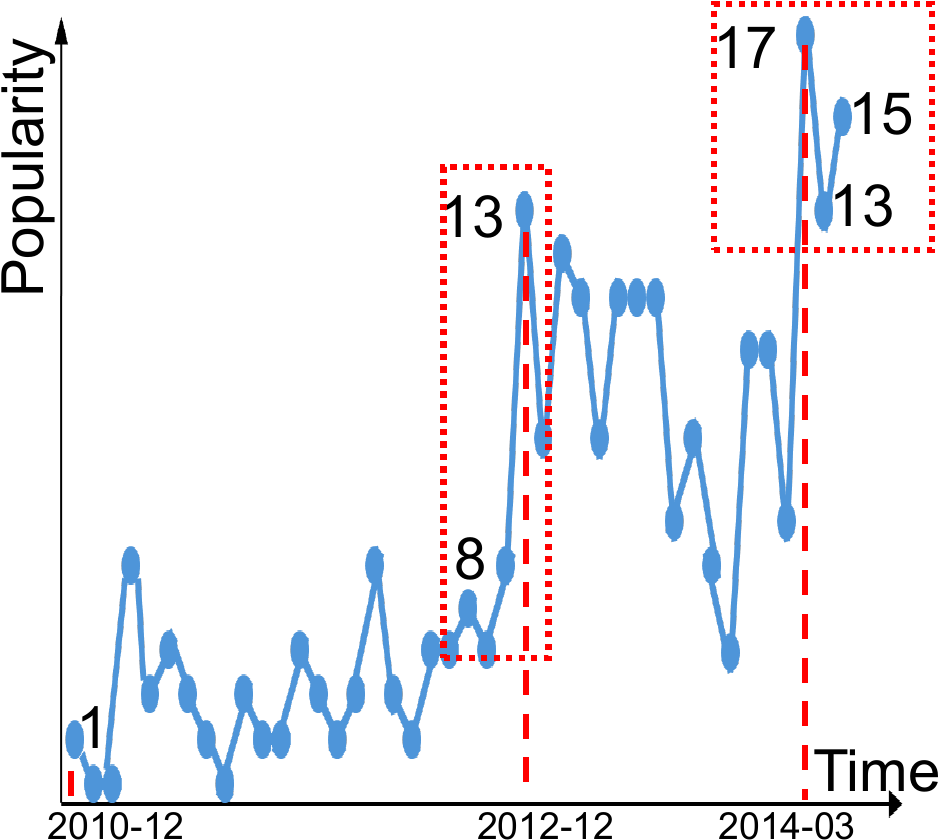}
    & 
    \textbf{Trend}: The historical popularity trends show a general upward trajectory over time, with \textcolor{red}{notable peaks in early 2013 and 2014}. The most recent peak in \textcolor{red}{March 2014 had a score of 17}, suggesting strong interest and positive reception of the product. \textbf{Feature}: The toothbrush's \textcolor{ForestGreen}{unique banana shape}, flexible design, and safety features (like being made of \textcolor{ForestGreen}{high-quality silicone, dishwasher safe, and freezer-friendly}) align well with current parental preferences for \textcolor{ForestGreen}{engaging, safe, and convenient infant products}. These aspects contribute to a higher popularity score. \textbf{Integration}: Combining the upward trend with the product's innovative features, which are likely to attract parents seeking effective yet appealing options, the predicted score for June 2014 is 15, reflecting a continuation of positive momentum.
    &
    15 \ \ (9)
    &
    0.6809 (6)
    &
    15 (2 - 7)
    \\
    \bottomrule
  \end{tabular}
  }
\vspace{-0.2cm}
\end{table*}

We further analyze the contribution of \shortname{}’s learning objectives by comparing the full model and two variants against the best baseline (PARE) and a ground-truth upper bound. The upper bound ranks items by their true test-set interaction counts, representing the best possible non-personalized recommendation. We denote w/o $\mathcal{L}_{\text{sup}}$ as training without the supervised prediction loss and w/o $\mathcal{L}_{\text{con}}$ as training without the contrastive loss, isolating the effect of each objective.


As shown in Table~\ref{tab:ablation}, \shortname{} outperforms both variants across all metrics, illustrating the benefits of combining supervised and contrastive objectives. The supervised loss contributes most directly to prediction accuracy, while the contrastive loss further improves both prediction and recommendation. We also observe that better popularity prediction generally translates to better recommendation, consistent with our assumption and prior work \cite{jing2023pare}. Finally, a clear gap remains to the ground-truth upper bound, especially in popularity-dominated domains. On \textit{Douban Movie}, the upper bound exceeds \shortname{} by 50.22\% in HR@10 and 40.07\% in NDCG@10.



\subsubsection{Cold-Start and Trend-Aware Evaluation}
To assess robustness under sparse supervision, we further evaluate TRAIL on Douban by partitioning test items into cold-start (fewer than 5 interactions) and warm-start groups. As shown in Figure \ref{fig:hr_compare_sasrec}, TRAIL delivers strong cold-start performance, where SASRec fails to produce any hits. TRAIL also maintains an advantage on warm-start items compared with SASRec, indicating consistent gains beyond cold-start regime.

In addition, we analyze performance across item popularity trend categories: increasing, decreasing, fluctuating, and unknown (0-1 record). Shown in Figure \ref{fig:hr_four_category}, TRAIL performs best on increasing items and remains competitive on decreasing and unknown groups, while fluctuating items are the most challenging. Notably, even under sparse or unstable trends, TRAIL exceeds the strongest baseline on full set, highlighting its adaptability in non-stationary settings.

\subsubsection{Explanation Performances}
To assess the quality of explanations generated by \shortname{}, we use an LLM-based evaluation over four dimensions: \textit{Trend Accuracy}, \textit{Metadata Faithfulness}, \textit{Predictive Consistency}, and \textit{Reasoning Coherence}. We apply standardized prompts and score each dimension on a 0–1 scale (prompt in Appendix), using GPT-4.1-mini as the judge to avoid circularity. Figure~\ref{fig:expain1} reports average results on \textit{Movie} and \textit{Baby}. We found \shortname{} consistently outperforms untrained LLM on all dimensions, with largest gains in \textit{Predictive Consistency} and \textit{Reasoning Coherence}, indicating more logically aligned and clearer explanations.

We conducted a human evaluation with ten participants. Each participant reviewed 20 randomly sampled explanation pairs, comparing GPT-generated explanations with those from \shortname{} and selecting the overall preferred one.
As shown in Figure~\ref{fig:expain2}, participants consistently preferred \shortname{}. On average, \shortname{} received over three times more votes than untrained LLM, indicating clearer, more informative, and more contextually accurate explanations. 

\subsubsection{Hyperparameter Sensitivity}
We conduct a hyperparameter study on the LoRA rank $r$ in Figure~\ref{fig:hyper}. Performance is largely stable across ranks, and $r=8$ yields the best results, indicating that a lower-rank adaptation is sufficient to capture task-specific knowledge effectively while maintaining efficiency.


\subsubsection{Case Study}
As shown in Table~\ref{tab:case}, we further conducted a case study using two examples from \textit{Douban Movie} and \textit{Amazon Baby}. Both examples highlight how \shortname{} integrates item features and temporal patterns to produce grounded popularity forecasts and coherent textual rationales. We illustrate more cases in Appendix.

For \textit{My Dear Liar (2019)} from \textit{Douban Movie}, where no historical data is available, \shortname{} correctly identifies the lack of trend information and bases its prediction on semantic cues. For the second \textit{Amazon Baby} product,
\shortname{} accurately captures the upward trend with peaks in 2013-2014 and links it to product features like safety, flexibility, and design. 
These examples show that \shortname{} combines content understanding with trend reasoning to produce accurate, well-grounded explanations. In contrast, PARE yields less accurate predictions and provides no textual explanations, highlighting \shortname{}’s advantages in both interpretability and prediction quality.


\section{Conclusion}
In this work, we present \shortname{}, a finetuned LLM that jointly predicts short-term item popularity and generates natural-language explanations for non-personalized recommendation. By leveraging item-side features and temporal popularity dynamics, \shortname{} provides scalable and interpretable recommendations without relying on exact user-item interactions. 
Experiments show that \shortname{} outperforms strong baselines in both popularity prediction and recommendation accuracy. Ablations and case studies further demonstrate robust reasoning on new items and emerging trends, producing clear, data-grounded explanations that traditional models lack, making \shortname{} well suited for large-scale deployment where efficiency, interpretability, and timeliness matter.

\begin{acks}
This research is supported by the RIE2025 Industry Alignment Fund – Industry Collaboration Projects (IAF-ICP) (Award I2301E0026), administered by A*STAR, as well as supported by Alibaba Group and NTU Singapore through Alibaba-NTU Global e-Sustainability CorpLab (ANGEL). This work is supported by Jinan-NTU Green Technology Research Institute (GreenTRI), and the Joint NTU-UBC Research Centre of Excellence in Active Living for the Elderly (LILY), Nanyang Technological University, Singapore.
\end{acks}


\appendix
\newpage
\section{Implementation Details}

In this section, we will present several important implementation details of \shortname{}, including the filtering of valid tokens during fine-tuning, the configuration of LoRA, and hyperparameters.

\subsection{Hyperparameters}

All models are implemented in PyTorch with the HuggingFace Transformers library. As we mentioned before, we choose DeepSeek-7B\footnote{https://huggingface.co/deepseek-ai/DeepSeek-R1-Distill-Qwen-7B} as the backbone of \shortname{}. And we applied LoRA to conduct the Parameter-Efficient Fine-Tuning. Table~\ref{tab:lora_config} summarizes the LoRA configurations explored in our study.

\begin{table}[ht]
\centering
\caption{LoRA configuration in training \shortname{}.}
\label{tab:lora_config}
\begin{tabular}{ccccc}
\toprule
\textbf{rank} & \textbf{lora\_alpha} & \textbf{target layer} & \textbf{dropout} & \textbf{bias} \\
\midrule
8 & 16 & \textit{q\_proj, v\_proj, o\_proj} & 0.05 & none \\
16 & 32 & \textit{q\_proj, v\_proj, o\_proj} & 0.05 & none \\
32 & 64 & \textit{q\_proj, v\_proj, o\_proj} & 0.05 & none \\
\bottomrule
\end{tabular}
\end{table}

Our main experiments were conducted with $r = 8$. For $r = 16$ and $r = 32$, and followed the usual setting, we set lora\_alpha to $2r$. Moreiver, we performed additional hyperparameter studies, and the corresponding results are presented in Figure \ref{fig:hyper}. We primarily fine-tuned the \textit{query, value, and output} layer, while all other layers (e.g., Feed-Forward Network) were frozen during training.

The setting above ensures that in the training process, LoRA adapters will be injected into all attention blocks of the transformer decoder, and only the projection matrices of the query, value, and output submodules were trainable, resulting in approximately $0.05\%$ to $0.1\%$ of the original parameters being updated during fine-tuning.

Given that we employ the InfoNCE loss, the temperature parameter $\tau$ serves as a key hyperparameter. Additionally, since the computation of InfoNCE requires measuring similarity between different training samples, we introduce three weighting coefficients $\alpha$, $\beta$, and $\gamma$ as additional hyperparameters. In our implementation, considering that historical trends and item metadata are the two most influential factors, we set $\alpha = \gamma = 0.4$ and $\beta = 0.2$. 

To enhance the discriminative power of the InfoNCE objective, the temperature $\tau$ is set to $0.1$. Due to the constraints of our training hardware, the batch size per GPU is set to $8$. For each anchor sample, we select the top-2 most similar samples as the positive set $S^{+}$, while treating the remaining samples as the negative set $S^{-}$. To enlarge the effective batch size and stabilize the training process, 
we set the gradient accumulation steps to $4$.

\subsection{Optimization \& Scheduler}

We used AdamW as the optimizer with a learning rate in $[1 \times 10^{-5}, 2 \times 10^{-5}]$. Moreover, we applied cosine decay scheduling. Additionally, we set the weight decay to 0.05 to prevent overfitting, adopted a cosine learning rate scheduler, and applied a $10\%$ warm-up ratio to ensure stable optimization in the early training stage.

Training was performed in \texttt{bf16} precision to enhance numerical stability, and gradient clipping was set to $1.0$ to prevent gradient explosion. 
Early stopping was applied with a patience of $3$, and model checkpoints were saved every $150$ steps. 
All experiments were conducted on NVIDIA H20 GPUs. And we fixed random seeds across all components for reproducibility.

\subsection{Sequence Processing for Contrastive Learning}

In computing the supervised loss, we only supervise the response portion of the output. To facilitate the later extraction of both the predicted score $p_i^T$ and the textual explanation of the predicted score, our implementation is designed to make the model output a JSON-formatted string. Accordingly, the supervised loss is applied specifically to the token span corresponding to the predicted score field. We follow the standard practice of using a mask value of $-100$ to exclude non-target tokens from the supervised loss computation.

Moreover, since we need to retrieve the embeddings associated with the explanation text, we introduce two special tokens, <exp\_start> and <exp\_end>, which allow us to easily locate and extract the tensor of explanation-related embeddings from the last hidden layer of LLM.

For input sequence processing, we perform padding based on the maximum sequence length within each batch, with an upper limit of 4096 tokens. Sequences longer than this limit are truncated from the end, while shorter ones are padded with the <pad> token. To guarantee valid computation of the contrastive loss, samples that do not contain both <exp\_start> and <exp\_end> tokens are deprecated during preprocessing.



\section{Prompt Design}

In our implementation of \shortname{}, multiple types of prompts are utilized to support different stages of the workflow. Specifically, we design distinct prompt templates for model training, inference, and LLM-based evaluation. The design of prompts is shown in Table \ref{tab:prompt_table}.

\begin{table*}[t]
  \caption{System Prompts and Their Usage}
  \centering
  \label{tab:prompt_table}
  \renewcommand{\arraystretch}{1.2}
  \resizebox{\textwidth}{!}{
  \small
  \begin{tabular}{
    >{\arraybackslash}m{0.95\linewidth}
    >{\centering\arraybackslash}m{0.1\linewidth}
  }
  \toprule
  \textbf{Prompt Text} & \textbf{Usage} \\
  \midrule
    \#\#\# Instruction: You are a movie popularity analyst. Your task is to predict a movie's popularity score based on its features and historical data, providing a structured explanation. Your output MUST be a single, valid JSON object with only two keys: predict\_popularity\_score and explanation\_of\_score.

    \#\#\# Output Requirements:
    1. predict\_popularity\_score: The predicted movie popularity score at the given ..., output a float. 2. explanation\_of\_score: A text with three mandatory sections, in order: [Trend]: Using concrete numbers (e.g., recent value, average, peaks/troughs, ...) [Feature]: Refer to at least two specific aspects from <MovieDescription> and explain ... [Integration]: Combine [Trend] and [Feature], and justify why the predicted score is reasonable at <InferenceTimestamp>.

    \#\#\# Inputs:
    
    <MovieDescription>\{ moviedescription \}</MovieDescription> 
    
    <PopularityHistory>\{ popularityhistory \}</PopularityHistory>
    
    <InferenceTimestamp>\{ inferencetimestamp \}</InferenceTimestamp>

    \#\#\# Answer:
    
    \{"predict\_popularity\_score": "xxx", "explanation\_of\_score": "<exp\_start>xxx<exp\_end>"\}
    & Training on Douban Movie dataset  \\
    \midrule
         \#\#\# Instruction: You are a movie popularity analyst. Your task is to predict a movie's popularity score ...

    \#\#\# Output Requirements:
    1. predict\_popularity\_score: T... 2. explanation\_of\_score: A text with three mandatory sections, in order: ...

    \#\#\# Inputs:
    
    <MovieDescription>\{ moviedescription \}</MovieDescription> 
    
    <PopularityHistory>\{ popularityhistory \}</PopularityHistory>
    
    <InferenceTimestamp>\{ inferencetimestamp \}</InferenceTimestamp>  & Inference on Douban Movie dataset  \\
    \midrule
    \#\#\# Instruction: 
    Your goal is to strictly and objectively assess the explanation's quality...  
    You will be given the true historical trend data, item description and metadata, the prediction timestamp, the ground-truth and predicted popularity scores, and the model’s generated explanation.  
    Please rate the explanation from 4 perspectives (scores between 0 and 1) and give a short comment:

    1. how the explanation describes the historical trend — does it accurately reflect the real trend’s direction, peaks, and fluctuations? Give a score between 0 and 1 for how truthful it is with respect to the historical trend, where 1 means completely accurate and 0 means completely inconsistent.

    2. metadata — are the mentioned product attributes (such as item name, item special feature, etc.) actually present well in the provided item description? Give a score between 0 and 1 for how truthful it is with respect to the metadata, where 1 means completely accurate” and 0 means completely inconsistent.

    3. Evaluate whether the explanation's reasoning about future popularity makes sense given the past trend. If the trend shows steady growth and the explanation predicts continued or strong demand, that’s consistent and should get a high score. If the explanation predicts something that clearly contradicts the trend without any justification, that’s inconsistent and should get a low score. Rate it from 0 to 1, where 1 means highly consistent and reasonable.
    
    4. Give a score from 0 to 1 based on how logical, consistent, and well-written the explanation is. Do not judge formatting or section structure; focus only on clarity, coherence, and internal logic.

    Return your output in JSON. & GPT Evaluation of generated explanation  \\

  \bottomrule
  \end{tabular}
  }
\end{table*}

\section{More Cases}
Table \ref{tab:case} presents additional examples illustrating \shortname{}’s interpretability and reasoning process.
The first case demonstrates a failure example where \shortname{} incorrectly predicts a high popularity score (16) for an item that actually received almost no interactions. This overestimation likely stems from overreliance on strong product features such as ergonomic design and target age suitability, despite weak historical signals. 
In contrast, the second case shows that \shortname{} accurately captures a declining trend, noting a brief popularity peak in March 2014 followed by a drop in subsequent months. Its explanation correctly connects this pattern with product functionality and brand appeal, leading to a moderate and reasonable score.
The third case, Joker (2019), highlights \shortname{}’s ability to link temporal and semantic cues. It identifies the sharp rise in popularity from September to October, aligning it with the film’s Venice Film Festival win and its high-profile cast, producing a strong and well-grounded prediction ranking close to the groundtruth.
Overall, these cases show that \shortname{} effectively combines trend and content reasoning to produce accurate, data-grounded explanations, though it may slightly overestimate popularity when strong feature cues lack trend support.

\begin{table*}[t]
  \caption{Additional Cases. For popularity, values are shown as $a(b)$, where $a$ is predicted score and $b$ is the corresponding rank.}
  \resizebox{\textwidth}{!}{
  \centering
  \small
  \label{tab:bad_case}
  \renewcommand{\arraystretch}{1.2}
  \begin{tabular}{
    >{\arraybackslash}m{0.3\linewidth} 
    >{\centering\arraybackslash}m{0.17\linewidth} 
    >{\arraybackslash}m{0.48\linewidth}
    >{\arraybackslash}m{0.1\linewidth}
    >{\arraybackslash}m{0.1\linewidth}
  }
    \toprule
    \textbf{Item Description} & 
    \textbf{Historical  Trend} & 
    \textbf{\shortname{} Explanation} &
    \textbf{\shortname{}} &
    \textbf{Groundtruth}
    \\
    \midrule
    The baby product name is Munchkin 6 Pack Soft-Tip Infant Spoon ..., The \textcolor{ForestGreen}{special handle is ergonomically designed} ... This is ideal for 3+ month's baby.; ... The category of the product are Baby; The price of the item is 5.09.
    & 2008-11: 1, 2008-12: 0, 2009-01: 0, ... 2013-02: 7, ... \textcolor{red}{2013-12: 16, 2014-01: 4, 2014-02: 10, 2014-03: 12}, 2014-04: 6, 2014-05: 9
    &  \textbf{[Trend]:} The historical popularity trends score \textcolor{red}{peaked at 16 in December 2013} and \textcolor{red}{continued to rise} in subsequent years. \textbf{[Feature]:} The product's features, such as being \textcolor{ForestGreen}{specifically designed for infants and ergonomic handle}, align with the target customer and contribute to popularity. \textbf{[Integration]:} Combining the upward trend ..., the predicted score is 16, reflecting continued growth.
    & 16 (7) & 1 ($\geq 856$)
    \\ \midrule
    The baby product name is {Philips AVENT} Drying Rack; The Philips dries your baby's bottles and ... The brand of this baby product is \textcolor{ForestGreen}{Philips Avert}; ..., The price of the item is 13.99. & 2014-01: 7, 2014-02: 11, \textcolor{red}{2014-03: 14}, 2014-04: 3, 2014-05: 2 & \textbf{Trend:} The historical popularity scores \textcolor{red}{show a peak in March 2014 with a score of 14}, followed by \textcolor{red}{a significant drop} in April and May, ... indicates a potential decline ... \textbf{Feature:} The product's capacity ..., along with its \textcolor{ForestGreen}{flexible design} ..., makes it highly functional and appealing for parents. As a product from a \textcolor{ForestGreen}{reputable brand}, ... benefits from brand loyalty. \textbf{Integration:} Combining the declining trend ..., the moderate score is fine. & 5.5 (194) & 1 ($\geq 856$)
    \\
    \midrule
    The movie Joker (I) (2019) is a Plot, Thriller ... \textcolor{ForestGreen}{This film won the Venice Film Festival} ... It is directed by \textcolor{ForestGreen}{Todd Phillips}, ... And the main actors (top-5) include … \textcolor{ForestGreen}{Robert De Niro}, Zazie Beetz, ..., Todd Phillips as producer, ... & 2019-08: 3, \textcolor{red}{2019-09: 13, 2019-10: 161} & \textbf{Trend:} The popularity score \textcolor{red}{increased significantly from 13 in September to 161 in October}, indicating strong growth momentum. The movie's win at the \textcolor{ForestGreen}{Venice Film Festival} and its \textcolor{ForestGreen}{high-profile cast} likely contributed to this trend. \textbf{Feature:} The movie's cast and production, including \textcolor{ForestGreen}{Todd Phillips} and \textcolor{ForestGreen}{Robert De Niro}, enhance its appeal ... \textbf{Integration:} Combining ..., a score of 170 in November is reasonable as it \textbf{follows the upward trend} ... & 
    170 (3) & 482 (1)
    \\
    \bottomrule
  \end{tabular}
  }
\end{table*}

\end{document}